\documentclass[12pt,journal,onecolumn]{IEEEtran}

\usepackage{threeparttable,placeins,makecell,stfloats,cite}
\usepackage{threeparttable,arydshln}
\usepackage{amsmath,amssymb,bm}
\usepackage[dvips]{graphicx}
\usepackage[usenames]{color}
\usepackage{amsfonts}
\usepackage{latexsym}
\usepackage{multirow}
\usepackage{caption}
\usepackage{rotating}
\usepackage{float}
\usepackage[format=hang]{subfig}
\usepackage{url}
\usepackage{cite}

\usepackage{booktabs}
\usepackage{amsmath}
\usepackage{algorithm,algorithmic}
\usepackage{tikz}
\usepackage{nicematrix}
\usetikzlibrary{arrows.meta,calc,positioning,quotes,fit}
\usepackage{stfloats}

\begin{document}
	\title{Novel Power-Imbalanced Dense Codebooks for Reliable Multiplexing in Nakagami Channels}
	\author{Yiming~Gui, Zilong~Liu, Lisu~Yu, Chunlei~Li,~and~Pingzhi~Fan
		\thanks{
    Y. Gui is with Xihua University, Chengdu 610039, China (e-mail: ymgui\_18@163.com).
    Z. Liu is with the School of Computer Science and Electronics Engineering, University of Essex, UK (e-mail: zilong.liu@essex.ac.uk). 
    L. Yu is with the School of Information Engineering, Nanchang University, Nanchang 330031, China (e-mails: lisuyu@ncu.edu.cn)
    C. Li is with the Department of Informatics, Selmer Center, University of Bergen, Norway (e-mail: chunlei.li@uib.no). 
    P. Fan is with The Key Lab of Information Coding and Transmission, Southwest Jiaotong University, Chengdu, China (e-mail: pzfan@swjtu.edu.cn). 
  }
	}
	\maketitle
	
\begin{abstract}
	This paper studies enhanced dense code multiple access (DCMA) system design for downlink transmission over the Nakagami-$m$ fading channels. By studying the DCMA pairwise error probability (PEP) in a Nakagami-$m$ channel, a novel design metric called minimum logarithmic sum distance (MLSD) is first derived. With respect to the proposed MLSD, we introduce a new family of power-imbalanced dense codebooks by deleting certain rows of a special non-unimodular circulant matrix. Simulation results demonstrate that our proposed dense codebooks lead to both larger minimum Euclidean distance and MLSD, thus yielding significant improvements of error performance over the existing sparse code multiple access and conventional unimodular DCMA schemes in Nakagami-$m$ fading channels under different overloading factors. 
\end{abstract}


\section{Introduction}
 
\IEEEPARstart{I}{n} recent years, with tens of billions of communication devices, sensors, robots, and vehicles connecting into the machine-type communication (MTC) networks, non-orthogonal multiple access (NOMA) has emerged as a disruptive technique for massive wireless access. There are two major types of NOMA: code-domain NOMA (CD-NOMA) which is a variant of the conventional code division multiple access \cite{Liu2021} and power-domain NOMA in which two or more users are separated by transmitting with different power levels \cite{Liu2017}.
	
In this paper, we focus on CD-NOMA which relies on good sequences or codebooks with certain distance properties. Thanks to the low-complexity {message passing algorithm (MPA)}, low-density spreading (LDS) systems were proposed in \cite{Hoshyar2008}. For larger constellation shaping gain, sparse code multiple access (SCMA) was developed in \cite{Nikopour2013,Taherzadeh2014} in which input bit messages are directly mapped to sparse codewords (unlike the direct spreading in LDS). In the literature, SCMA codebooks can be efficiently generated by applying different operations (such as interleaving, permutation, phase rotations) to a common multi-dimensional mother constellation \cite{Chaturvedi2022,Yu2018,Chen2020,Mheich2018,Li2022,Wen2022,Luo2022}.
	
Numerous research attempts have been made on SCMA in the past years. However, its bit error rate (BER) performance over a Rayleigh fading channel suffers from a small diversity order that is equal to the effective number of resource elements occupied by each user \cite{Lim2017}. In contrast, dense code multiple access (DCMA) owns the potential of full diversity order and hence improved BER performance over fading channels, as pointed out in \cite{Liu2021}. Such full diversity can be acquired by adopting a non-linear sphere decoder at the receiver. 

Up to date, however, DCMA codebook optimization remains largely open. An important research problem is the design of dense codebooks which exhibit good adaptability to various fading channels. Such a design is of strong interest to massive connectivity in, for example, a \textit{non-stationary} vehicular communication channel, where moving connected vehicles may experience varying channel coefficients \cite{Md2022}. Motivated by this, we aim for developing enhanced dense codebooks for reliable downlink MTC over the Nakagami-$m$ fading channels that are well suited for both {line-of-sight (LOS) and non-LOS transmissions} in vehicular communications \cite{He2014}.
	
First, we derive the pairwise error probability (PEP) of DCMA over the generalized Nakagami-$m$ fading channel, where the dense codebooks are optimized with respect to a novel minimum logarithmic sum distance (MLSD), as shown in (\ref{MLSD}) of Section III. We further show that the PEPs for Gaussian and Rayleigh fading channels can be deduced as two special cases of that for generalized Nakagami-$m$ fading channel for $m=+\infty$ and $m=1$, respectively. Unlike the existing unimodular DCMA codebooks in \cite{Liu2021}, we propose a novel dense codebook design by properly deleting certain rows of a non-unimodular circulant matrix. On one hand, the non-unimodular codebooks yield certain power imbalance among different users which is the key for amplifying the user separation. On the other hand, the proposed circulant matrix structure significantly reduces the design complexity as one only needs to optimize $J+1$ parameters, where $J$ refers to the number of users. Numerical simulations show that our proposed power-imbalanced dense codebooks lead to both larger minimum Euclidean distance (MED) and MLSD and hence significantly improved BER performance compared to existing SCMA and unimodular DCMA.

\section{System Model}
\subsection{System Model} 	
We consider a downlink DCMA system for downlink MTC, where a base station (BS) communicates with $J$ user equipments (UEs) sharing $K$ resource elements (REs), where $J>K$ for overloading factor $\lambda=J/K>1$. Both the BS and UEs are equipped with a single antenna. At the BS side, each user's codeword $\bm{x}_j$ can be obtained by multiplying the quadrature amplitude modulation (QAM) symbol $\alpha_j \in \mathbb{C}$ with a distinctive sequence $\bm{d}_j\in \mathbb{C}^{K\times 1}$. Thus, the received codeword at user $j$ can be written as follows:
	\begin{equation}
	\begin{split}
		\bm{y}_j&=\text{diag}\left(\bm{h}_j\right)\sum_{j=1}^{J}\alpha_j\bm{d}_j+\bm{n}_j\\
		&=\text{diag}\left(\bm{h}_j\right)\bm{\mathcal{D}}\bm{\alpha}+\bm{n}_j=\bm{G}\bm{\alpha}+\bm{n}_j,
	\end{split}
\end{equation}
where $\bm{\alpha}=\left[\alpha_1,\cdots,\alpha_J\right]^T$ represents the transmit symbol vector of $J$ users and $\left(\cdot\right)^T$ denotes the transpose operation. $\bm{\mathcal{D}}=\left[\bm{d}_1,\cdots,\bm{d}_J\right]$ is a dense signature matrix with no zero entries. $\bm{h}_j=\left[h_{1,j},\cdots,h_{K,j}\right]^T$ denotes the channel fading vector. Furthermore, $\bm{n}_j=\left[n_{1,j},\cdots,n_{K,j}\right]^T$ stands for the Gaussian noise with $n_{k,j}\sim \mathcal{CN}\left(0,N_0\right)$. In this work, we consider the typical Nakagami-$m$ fading channel for vehicular communications \cite{He2014, Belmekki2020}. 
\subsection{Detection Algorithm} 	
Sphere decoder (SD) is widely adopted to solve linear multiple-input and multiple-output (MIMO) equations. Similar to \cite{Liu2021}, we utilize the generalized sphere decoder (GSD) proposed in \cite{Cui2004, Luo2021} to tackle the rank-deficient linear equations resulting from $K<J$.  The algorithm for GSD detection can be expressed as follows:
\begin{equation}
	\label{GSD}
	\begin{split}
		\hat{\bm{\alpha}}&=\text{arg}\min_{\bm{\alpha}}\left(\left\|\bm{y}-\bm{G}\bm{\alpha}  \right\|^2+\lambda \bm{\alpha}^{\mathcal{H}}\bm{\alpha}\right) \\
		&=\text{arg}\min_{\bm{\alpha}}\left(\bm{y}^{\mathcal{H}}\bm{y}-\bm{y}^{\mathcal{H}}\bm{G}\bm{\alpha}-\bm{\alpha}^{\mathcal{H}}\bm{G}^{\mathcal{H}}\bm{y}+\bm{\alpha}^{\mathcal{H}}\bm{Q}\bm{\alpha}\right) \\
		&=\text{arg}\min_{\bm{\alpha}} \left\| \bm{r}-\bm{U}\bm{\alpha} \right\|,
	\end{split}
\end{equation}
where $\left(\cdot\right)^{\mathcal{H}}$ indicates Hermitian transpose operation, $\lambda$ denotes a positive constant, $\bm{Q}=\bm{G}^{\mathcal{H}}\bm{G}+\lambda \bm{I}_J $ is the positive definitive matrix and can be decomposed as  $\bm{Q}=\bm{U}^{\mathcal{H}}\bm{U}$, where $\bm{U}$ denotes the upper triangular matrix.

According to (\ref{GSD}), the linear equation system can be converted to a full-rank one. Therefore, a conventional SD algorithm may be utilized to detect the transmitting signals.

{Unlike the GSD algorithm adopted in DCMA, the multiuser detection in SCMA is carried out with the aid of MPA by efficiently exploiting the codebook sparsity. As discussed in \cite{Liu2021}, the complexity of GSD is approximately 35\% and 133\% of that of MPA in the $(4\times 6)$ and $(5\times 10)$ settings, respectively.}

\section{Proposed Dense Codebook Design}
\subsection{Proposed Design Criteria}
Let us define the element-wise distance $\delta_{\bm{w}\to \bm{\hat{w}}}\left(k\right)=\left(\bm{\mathcal{D}}\left(\bm{\alpha}-\bm{\hat{\alpha}}\right)\right)_k$, where $\bm{w}=\bm{\mathcal{D}}\bm{\alpha}$. The PEP conditioned on channel vector $\bm{h}$ between $\bm{w}$ and $\bm{\hat{w}}$ can be written as follows:
\begin{equation}
	\label{PEP-Condition}	
	\text{Pr}\left(\bm{w}\to \bm{\hat{w}}\right | \bm{h})=Q\left(\sqrt{\frac{\sum_{k=1}^{K}|h_k|^2|\delta_{\bm{w}\to \bm{\hat{w}}}\left(k\right)|^2}{2N_0}}\right),
\end{equation}
where $Q\left(x\right)=\left(2\pi\right)^{-1/2}\int_{x}^{+\infty}e^{-t^2/2}\text{d}t$ denotes the Gaussian Q-function. For a sufficiently high signal-to-noise (SNR) region, we propose to use curve fitting to approximate the Q-function for $x\geq 1.5$. In this case, it is found that 
\begin{equation}
	\label{Qfunction-Approximation}
	Q\left(x\right)\simeq \frac{3}{10}e^{-\frac{3x^2}{5}}.
\end{equation}

Applying (\ref{Qfunction-Approximation}) into (\ref{PEP-Condition}), we obtain
\begin{equation}
	\label{PEP-Condition-Approximation}
		\text{Pr}\left(\bm{w}\to \bm{\hat{w}}\right | \bm{h}) \simeq  
		\frac{3}{10}\prod_{k=1}^{K}\text{exp}\left(-\frac{3|\delta_{\bm{w}\to \bm{\hat{w}}}\left(k\right)|^2}{10N_0}|h_k|^2\right).
\end{equation}

The probability density function (pdf) of the Nakagami-$m$ channel model can be shown as follows:
\begin{equation}
	f\left(x\right)=\frac{2m^m}{\Gamma\left(m\right)\Omega^m}x^{2m-1}e^{-\frac{m}{\Omega}x^2},
\end{equation}
where $\Omega$ and $\Gamma\left(\cdot\right)$ represent the average transmitting  power and Gamma function, respectively. $m$ denotes the fading parameter. When $m=1$, the Nakagami-$m$ model reduces to the Rayleigh fading model, whilst $m=+\infty$ leads to a Gaussian channel. Thus, the pdf of power $z=x^2$, which follows a Gamma distribution, may be obtained as follows:
\begin{equation}
	f\left(z\right) = \frac{m^m}{\Gamma\left(m\right)\Omega^m}z^{m-1}e^{-\frac{m}{\Omega}z}.
\end{equation}

To proceed, we derive the moment generating function of  $|h_k|^2$ as follows:
\begin{equation}
	\label{MGF}
	\begin{split}
		&M_{z}\left(s\right)=E\left(e^{-sz}\right)\!=\!\int_{0}^{+\infty}\!\frac{m^m}{\Gamma\left(m\right)\Omega^m}z^{m-1}e^{-\tilde{s}z}\text{d}z \\
		&=\int_{0}^{+\infty}\frac{m^m}{\Gamma\left(m\right)\left(m+s\Omega\right)^m}\left(\tilde{s}z\right)^{m-1}e^{-\tilde{s}z}\text{d}\tilde{s}z \\
		&=\frac{1}{\left(1+\frac{s\Omega}{m}\right)^m},
	\end{split}
\end{equation}
where $\tilde{s}=\left(m/\Omega+s\right)$. 

Without loss of generality, we set the average transmitting power $\Omega$ to 1. By taking expectation to (\ref{PEP-Condition-Approximation}) with regard to $\bm{h}$, we arrive at
\begin{equation}
	\label{PEP-Uncondition}
		\text{Pr}\left(\bm{w}\to \bm{\hat{w}}\right) \simeq 
		\frac{3}{10}\prod_{k=1}^{K}\left(1+\frac{3|\delta_{\bm{w}\to \bm{\hat{w}}}\left(k\right)|^2}{10mN_0}\right)^{-m}.
\end{equation}

We next consider some special cases of (\ref{PEP-Uncondition}).

\emph{Case 1:} For $m\gg \frac{3|\delta_{\bm{w}\to \bm{\hat{w}}}\left(k\right)|^2}{10N_0}$ and thus $\frac{3|\delta_{\bm{w}\to \bm{\hat{w}}}\left(k\right)|^2}{10mN_0}\to 0$, we have $\left(1+\frac{3|\delta_{\bm{w}\to \bm{\hat{w}}}\left(k\right)|^2}{10mN_0}\right)^{-m}\approx \text{exp}\left(-\frac{3|\delta_{\bm{w}\to \bm{\hat{w}}}\left(k\right)|^2}{10N_0}\right)$. In this special case, (\ref{PEP-Uncondition}) reduces to
\begin{equation}
	\text{Pr}\left(\bm{w}\to \bm{\hat{w}}\right)\simeq \frac{3}{10}\text{exp}\left(-\frac{3\sum_{k=1}^{K}|\delta_{\bm{w}\to \bm{\hat{w}}}\left(k\right)|^2}{10N_0}\right),
\end{equation}
where the Euclidean distance $\sum_{k=1}^{K}|\delta_{\bm{w}\to \bm{\hat{w}}}\left(k\right)|^2$ dominates the PEP and this is consistent with {fact that the MED is a key performance metric for Gaussian channel with $m\to +\infty$. Here, the MED $\Delta_{\min}$ can be expressed as
\begin{equation}
    \Delta_{\min}=\min_{\bm{w}\neq \bm{\hat{w}}}\sum_{k=1}^{K}\delta_{\bm{w}\to \bm{\hat{w}}}\left(k\right)^2.
\end{equation}
}

\emph{Case 2:} For $m\ll \frac{|\delta_{\bm{w}\to \bm{\hat{w}}}\left(k\right)|^2}{N_0}$ in high SNR region, we have $\frac{|\delta_{\bm{w}\to \bm{\hat{w}}}\left(k\right)|^2}{mN_0}\gg 1$ and (\ref{PEP-Uncondition}) may reduce to
\begin{equation}
		\text{Pr}\left(\bm{w}\!\to\!\bm{\hat{w}}\right)\!\simeq \!
		\frac{10^{mK-1}\left(mN_0\right)^{mK}}{3^{mK-1}}\!\left(\!\prod_{k=1}^{K}\!|\!\delta_{\bm{w}\to \bm{\hat{w}}}\left(k\right)\!|\right)^{-2m},
\end{equation}
where the product distance $\prod_{k=1}^{K}|\delta_{\bm{w}\to \bm{\hat{w}}}\left(k\right)|$ dominates the PEP. Similarly, this is consistent with the results for Rayleigh fading channel with $m=1$.

From (\ref{PEP-Uncondition}), we obtain the proposed MLSD $\Xi_{\text{min}}$ of the superimposed constellation under Nakagami-$m$ channel below as a performance metric for the construction of enhanced dense codebooks:
\begin{equation}
	\label{MLSD}
	\begin{split}
		\Xi_{\text{min}}\triangleq \min_{\bm{w}\neq \bm{\hat{w}}} m\sum_{k=1}^{K}\text{log}\left(1+\frac{3|\delta_{\bm{w}\to \bm{\hat{w}}}\left(k\right)|^2}{10mN_0}\right),
	\end{split}
\end{equation}
where $\Xi_{\text{min}}$ depends on fading parameter $m$ and noise variance $N_0$.

\subsection{Proposed Circulant Codebook Design}
For minimum PEP, we are committed to developing DCMA codebooks for enhanced BER performance in the Nakagami-$m$ fading channel. We begin by presenting an $n\times n$ non-unimodular circulant matrix $\bm{\mathcal{C}}$ which is shown as follows:
\begin{equation}
	\label{circulant matrix}
	\bm{\mathcal{C}}=\frac{1}{\nu_n}\begin{bmatrix}
		c_{0} & c_{1} & \cdots & c_{n-1} \\
		c_{1} & c_{2} & \cdots & c_{0}  \\
		\vdots & \vdots & \ddots & \cdots \\
		c_{n-2} & c_{n-1} &\cdots & c_{n-3} \\
		c_{n-1} & c_{0} & \cdots & c_{n-2}
	\end{bmatrix},
\end{equation}
where $c_{k}=t^{\sqrt{k}}e^{i\varphi_{k}},k=0,\cdots,n-1$. $t$ is a positive number that effectively controls the distribution of energy for amplifying power variation among different users over the same resource node. $\left[\varphi_0, \varphi_1, \cdots,\varphi_{n-1}\right]$ denotes the phase argument vector which needs to be optimized for increasing the distance among superimposed constellation points and thus improving the BER performance. In addition, the square-root operation $\sqrt{\cdot}$ can effectively prevent $t^{\sqrt{n-1}}$ from rising sharply as the number of users increases, whereas $\nu_n$ is introduced to meet the power constraints. 

This paper adopts a circular structure to construct the signature matrix for the following reasons. Firstly, the circular matrix naturally satisfies the Latin structure, i.e., each distinct element appears only once in the same row or the same column. Different users in the same row have different magnitudes and phases. Through optimization, such a large design degree of freedom is useful for increasing the minimum distance among superimposed constellation points. Secondly, the circulant structure can significantly reduce the number of design parameters from $KJ$ to $J+1$.
	
For overloaded DCMA transmission, we propose to delete $J-K$ rows from a $J\times J$ circulant matrix. It is noted that row deletion leads to power imbalance among different users due to the inherent circulant matrix structure. For a large codebook distance, these deletion rows are judiciously selected by ensuring the minimum energy variation for the columns of the resultant $K\times J$ dense matrix. In this work, dense  matrices for $\bm{\mathcal{D}}_{4\times 6}$ and $\bm{\mathcal{D}}_{5\times 10}$ are obtained by setting the deletion rows of $\bm{\mathcal{C}}_{6\times 6}$ and $\bm{\mathcal{C}}_{10\times 10}$ as $\left[3,6\right]$ and $[2,4,6,8,10]$, respectively. 
	
As stated in the previous subsection, it is essential to maximize the MLSD in order to improve the BER performance of DCMA over Nakagami-$m$ fading channel. Let $\bm{\gamma}=\left(t,\varphi_{0},\cdots,\varphi_{J-1}\right)$ denote the set of parameters. The optimization of circulant matrix design may thus be formulated as follows: 
\begin{subequations}
	\label{Optimization}
	\begin{alignat}{4}
		\label{opt}
		\max_{\bm{\gamma}\in \mathbb{R}^{\left(J+1\right)\times 1}}&  \Xi_{\text{min}}\\
		\mbox{s.t.}\quad
		& tr\left(\bm{\mathcal{D}}^{\mathcal{H}}\bm{\mathcal{D}}\right)=J, \\
		& 0\leq t\leq 2,\\
		& 0\leq \varphi_j \leq \pi,\ j=0,\cdots,J-1,
	\end{alignat}
\end{subequations}
where $tr\left(\bm{\mathcal{D}}^{\mathcal{H}}\bm{\mathcal{D}}\right)\!=\!J$ confines the total power. 

It is quite difficult to directly solve the problem (\ref{Optimization}) since the objective function is non-convex. Therefore, \emph{fmincon} solver of  MATLAB Global Optimization toolbox is adopted in the work to obtain good MLSDs and MEDs which significantly outperform existing SCMA and unimodular DCMA. Through optimization, the obtained optimization results for $\bm{\mathcal{D}}_{4\times 6}$ and $\bm{\mathcal{D}}_{5\times 10}$ are $\bm{c}=(0.1887+0.0000i, 0.1894+0.2836i, 0.2245+0.3734i, -0.3983+0.3433i, -0.6145+0.0457i, 0.6753+0.2146i)$ and $\bm{c}=(1.0979+0.0000i, 0.1319+0.5265i, 0.0845+0.3965i,-0.2723+0.1756i, 0.2675+0.0212i, 0.0054+0.2271i,-0.1644+0.1057i,0.1623+0.0512i,-0.0782+0.1276i,-0.1307+0.0226i)$, respectively. 

\begin{figure}[!t]
	\centering
	\includegraphics[width=3in]{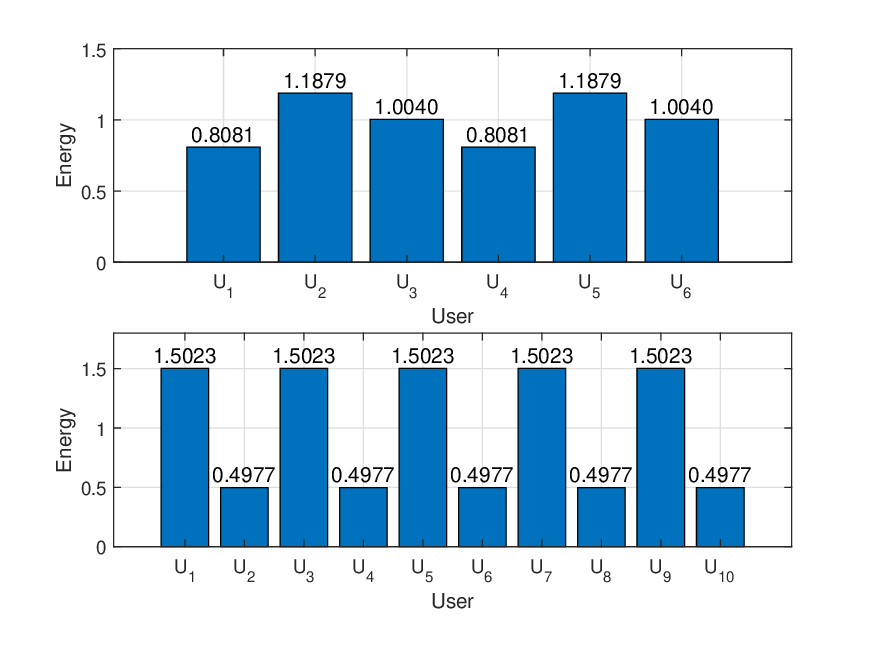} \\
	\caption{Energy of different users for $\bm{\mathcal{D}}_{4\times 6}$ and $\bm{\mathcal{D}}_{5\times 10}$.}
	\label{powers}
\end{figure}

It can be seen in Fig. \ref{powers} that row deletion leads to power imbalance among different users due to the inherent circulant matrix structure. For non-unimodular DCMA with $K=4$ and $J=6$, the maximum and minimum energy for different users are 1.1879 and 0.8081, respectively. For non-unimodular DCMA with $K=5$ and $J=10$, the maximum and minimum energy for different users are 1.5023 and 0.4977, respectively. 

\section{Numerical Results}
Table \ref{Comparison of different codebooks} compares the MLSDs and MEDs of the proposed dense codebooks, Yu's sparse codebook \cite{Yu2018}, Chen's sparse codebook \cite{Chen2020}, GAM sparse codebook \cite{Mheich2018} and Liu's unimodular DCMA \cite{Liu2021}. $m=4$ and $N_0=0.001$ was set in the codebook optimization as it leads to large MLSD. Clearly, our proposed DCMA enjoys larger MLSDs and MEDs. For $(4\times 6)$-DCMA system with overloading factor $\lambda=150\%$, the proposed scheme yields an MLSD and an MED of 40.20 and 1.24, respectively. For the $(5\times 10)$-DCMA with overloading factor of $200\%$, the proposed scheme yields an MLSD and an MED of 38.26 and 1.00, respectively.

\begin{figure}[!t]
	\centering
	\includegraphics[width=3.2in]{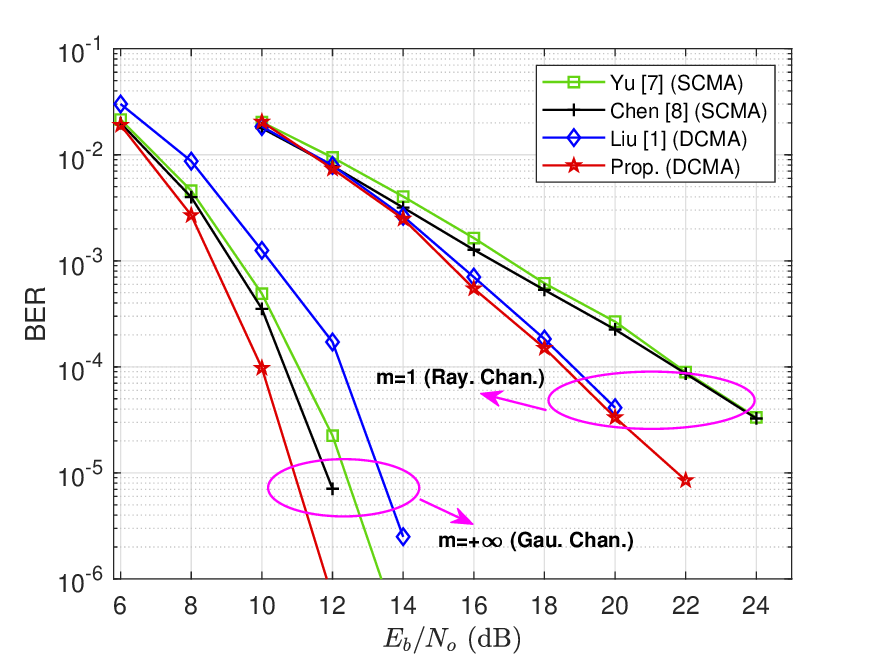} \\
	\caption{BER performance comparison under Gaussian and Rayleigh fading channels for $K=4$, $J=6$ and $\lambda=150\%$.}
	\label{BER_M4F4x6}
\end{figure}

In Fig. \ref{BER_M4F4x6}, we consider Gaussian channel and Rayleigh fading channel which are two special cases of the Nakagami-$m$ fading channel. As shown in Fig. \ref{BER_M4F4x6}, the proposed scheme outperforms that of the Yu's and Chen's codebooks under the Gaussian channel with $m=+\infty$. In particular, the proposed scheme achieves 1.7 dB and 2.6 dB gains over Yu's codebook and Liu's unimodular dense codebook at BER=$10^{-5}$, respectively. In downlink Rayleigh fading channel with $m=1$, the proposed scheme also leads to significantly improved error rate performance over Yu's and Chen's codebooks. In particular, the proposed DCMA yields 3.2 dB over Yu's codebook at BER=$10^{-4}$. 
\begin{table}[!t]
	\centering
	\caption{Comparison of Different Codebooks}
	\begin{tabular}{c  c c c c}
		\hline\toprule
		Overloading& Codebook  & MED & MLSD & Decoding Method\\ 
		\hline
		\multirow{4}*{$\lambda$=150\%}	 & Yu \cite{Yu2018} & 0.90 &28.38& MPA \\ 
		& Chen \cite{Chen2020} &1.07 &31.81& MPA\\ 
		& Liu \cite{Liu2021} & 0.78  &37.57& GSD\\
		& Prop. & 1.24 &40.20& GSD\\
		\hline
		\multirow{4}*{$\lambda$=200\%} & Yu \cite{Yu2018} & 0.48 &22.84& MPA\\
		& GAM \cite{Mheich2018} &0.43 &21.86& MPA\\	
		& Liu \cite{Liu2021} & 0.63  &32.22& GSD\\
		& Prop. & 1.00 &38.26& GSD\\
		\bottomrule\hline
	\end{tabular}
	\label{Comparison of different codebooks}
\end{table}

\begin{figure}[!t]
	\centering
	\includegraphics[width=3.2in]{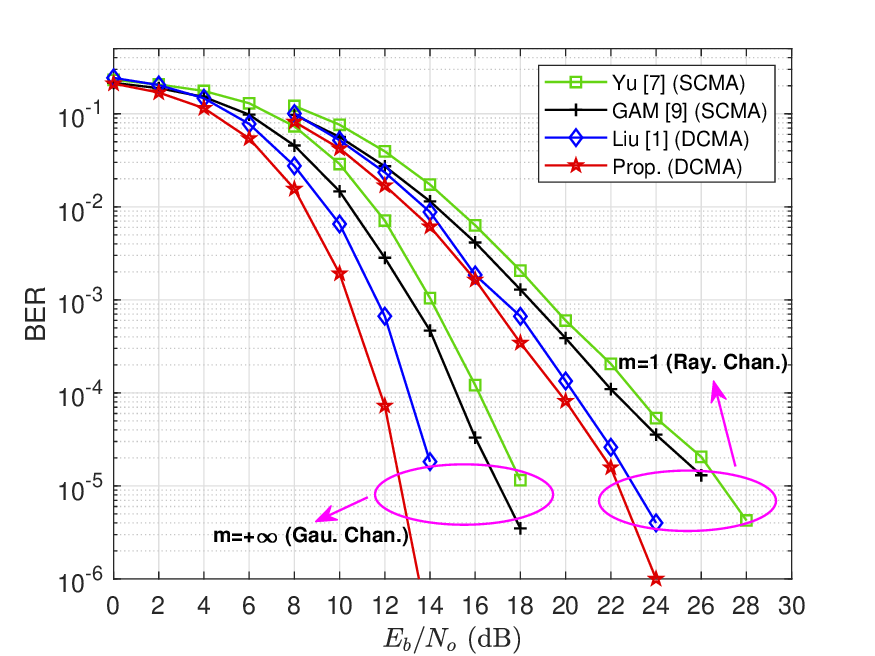} \\
	\caption{BER performance comparison under Gaussian and Rayleigh fading channels for $K=5$, $J=10$ and $\lambda=200\%$.}
	\label{BER_M4F5x10}
\end{figure}
\begin{figure}[!t]
	\centering
	\includegraphics[width=3.2in]{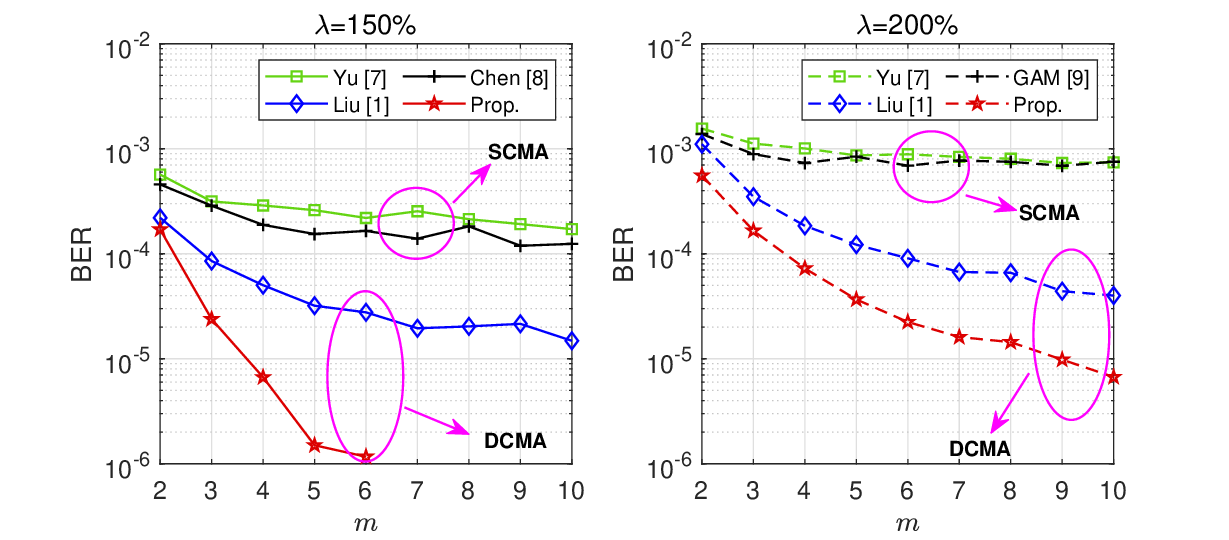} \\
	\caption{BER performance comparison under Nakagami-$m$ channel with different $m$ at $E_b/N_o=14$ dB.}
	\label{BER_Nakagami}
\end{figure}
\begin{figure}[!t]
	\centering
	\includegraphics[width=3.2in]{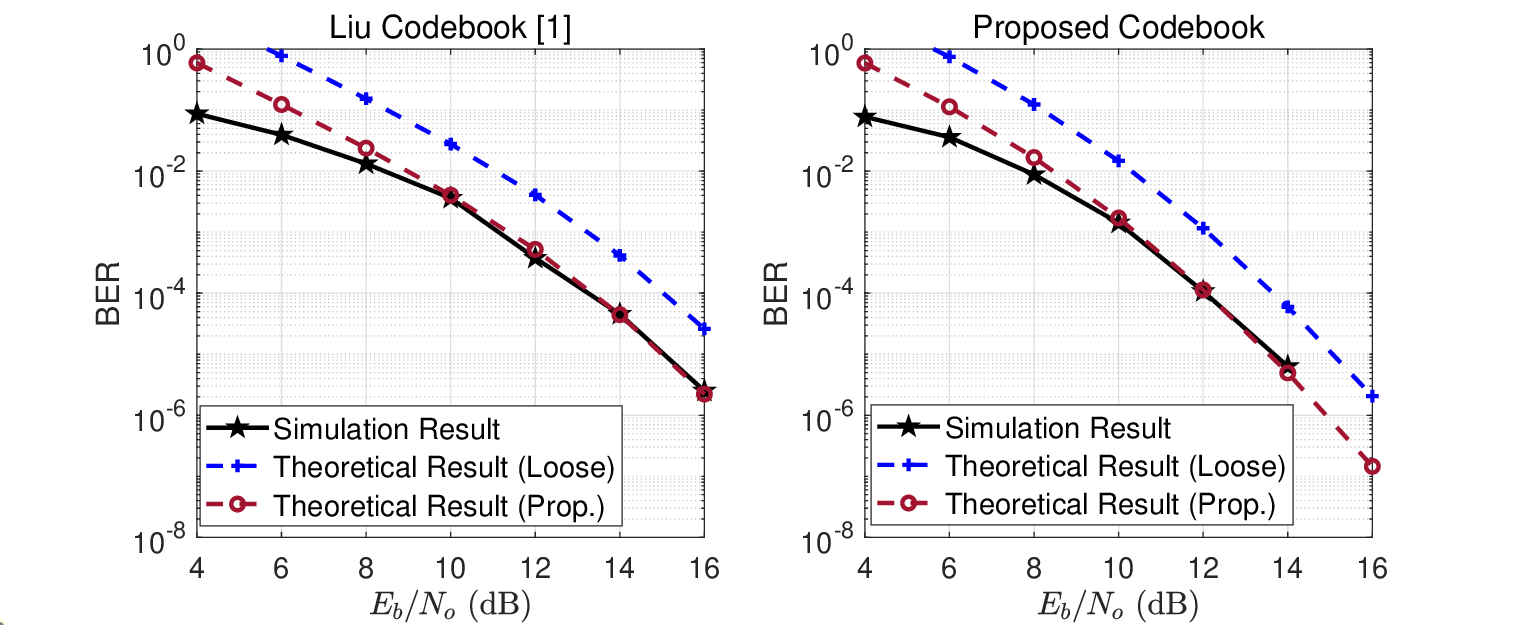} \\
	\caption{Comparison of theoretical and simulation results of BER over Nakagami-$m$ channel with $m=4$.}
	\label{simulation_theory_F4x6}
\end{figure}

In Fig. \ref{BER_M4F5x10}, we evaluate the BER performances of the proposed DCMA with $K=5$, $J=10$ and $\lambda=200\%$. We consider GAM codebook for comparison, which performs well in Gaussian and Rayleigh fading channels with $\lambda=200\%$. As shown in Fig. \ref{BER_M4F5x10}, our proposed scheme achieves the best BER performance in Gaussian and Rayleigh fading channels. In particular, the proposed scheme achieves nearly 1.6 dB and 4.3 dB gains over Liu's unimodular DCMA and Yu's codebook at BER=$10^{-5}$ under the Gaussian channel, respectively. In downlink Rayleigh fading channel, the proposed scheme leads to significantly improved error rate performance over Yu's and Chen's codebooks. In particular, the proposed DCMA using a circulant matrix yields 3.6 dB over Yu's codebook at BER=$10^{-5}$.

In Fig. \ref{BER_Nakagami}, we compare the BER performances of various codebooks in Nakagami-$m$ fading channels with different parameters $m$ at $E_b/N_o=14$ dB. With the increasing of $m$, the communication quality of Nakagami-m fading channel keeps improving. In this case, SCMA suffers from BER saturation when $m$ is 4 or greater, indicating a poor adaptability to different channel conditions. By contrast, DCMA generally exhibits improved BER for larger value of $m$, where the best BER improvement is achieved by our proposed dense codebooks.

To validate the our proposed PEP criteria for Nakagami-$m$ channel, we compare the simulation results and theoretical bounds. For the BER, the theoretical bound can be obtained by taking the average over all possible transmission symbols by the union bound, which can be written as follows:
\begin{equation}
	P_{b}\leq \frac{1}{M^J\cdot J\text{log}_2{M}}\sum_{\bm{w}}\sum_{\bm{\hat{w}}}n_{\text{E}}\left(\bm{w},\bm{\hat{w}}\right)\text{Pr}\left(\bm{w}\to \bm{\hat{w}}\right),
\end{equation}
where $n_{\text{E}}\left(\bm{w},\bm{\hat{w}}\right)$ denotes the number of erroneous bits when $\bm{\hat{w}}$ is decoded at the receiver. $\text{Pr}\left(\bm{w}\to \bm{\hat{w}}\right)$ is the PEP derived in (\ref{PEP-Uncondition}).

In Fig. \ref{simulation_theory_F4x6}, it is shown that the simulation results are well matched to theoretical bounds in high SNR region regardless of the Liu's dense codebook \cite{Liu2021} or the proposed dense codebook for Nakagami-$m$ channel with $m=4$. In addition, the loose theoretical results derived from $Q\left(x\right)\leq e^{-x^2/2}$ departs far from the simulation results and the proposed theoretical bound, which demonstrates that the proposed bound is tight for dense codebooks under the Nakagami-$m$ channel.

\section{Conclusions}
In this work, we have conceived a novel non-unimodular circulant DCMA codebook design and derived a new design criterion MLSD for Nakagami-$m$ channel. Compared to existing unimodular DCMA, the proposed scheme inherently introduces certain power imbalance among different users. As a result, larger MLSD can be achieved, leading to significantly improved BER performances over the existing SCMA and unimodular DCMA in Nakagami-$m$ fading channels with different $m$ values and overloading factors of $\lambda=150\%$ and $\lambda=200\%$.


\begin{thebibliography}{l}	

    \bibitem{Liu2021}
	Z. Liu and L. -L. Yang, ``Sparse or dense: a comparative study of code-domain NOMA systems," \emph{IEEE Transactions on Wireless Communications}, vol. 20, no. 8, pp. 4768-4780, Aug. 2021.
	
	\bibitem{Liu2017}
	Y. Liu, Z. Qin, M. Elkashlan, Z. Ding, A. Nallanathan, and L. Hanzo, ``Nonorthogonal multiple access for 5G and beyond,"
	\emph{Proceedings of the IEEE}, vol. 105, no. 12, pp. 2347-2381, Dec. 2017.
	
	\bibitem{Hoshyar2008}
	R. Hoshyar, F. P. Wathan, and R. Tafazolli, ``Novel low-density signature for synchronous CDMA systems over AWGN channel," \emph{IEEE Transactions on Signal Processing}, vol. 56, no. 4, pp. 1616-1626, Apr. 2008.
	
	\bibitem{Nikopour2013}
	H. Nikopour and H. Baligh, ``Sparse code multiple access," \emph{IEEE 24th Annual International Symposium on Personal Indoor and Mobile Radio Communications (PIMRC)}, 2013, pp. 332-336.

	\bibitem{Taherzadeh2014}
	M. Taherzadeh, H. Nikopour, A. Bayesteh, and H. Baligh, ``SCMA codebook design," \emph{IEEE 80th Vehicular Technology Conference (VTC2014-Fall)}, 2014, pp. 1-5.

       \bibitem{Chaturvedi2022}
	S. Chaturvedi, Z. Liu, V. Bohara, A. Srivastawa, and P. Xiao, ``A tutorial on decoding techniques of sparse code multiple access," IEEE Access, vol. 10, pp.  58503-58524, May 2022.

	\bibitem{Yu2018}
	L. Yu, P. Fan, D. Cai, and Z. Ma, ``Design and analysis of SCMA codebook based on Star-QAM signaling constellations," \emph{IEEE Transactions on Vehicular Technology}, vol. 67, no. 11, pp. 10543-10553, Nov. 2018.
	
	\bibitem{Chen2020}
	Y.-M. Chen and J.-W. Chen, ``On the design of near-optimal sparse code multiple access codebooks," \emph{IEEE Transactions on Communications}, vol. 68, no. 5, pp. 2950-2962, May. 2020.
	
	\bibitem{Mheich2018}
	Z. Mheich, L. Wen, P. Xiao, and A. Maaref, ``Design of SCMA codebooks based on golden angle modulation," \emph{IEEE Transactions on Vehicular Technology}, vol. 68, no. 2, pp. 1501-1509, Feb. 2019.
	
	\bibitem{Li2022}
	X. Li, Z. Gao, Y. Gui, Z. Liu, P. Xiao, and L. Yu, ``Design of power-imbalanced SCMA codebook," \emph{IEEE Transactions on Vehicular Technology}, vol. 71, no. 2, pp. 2140-2145, Feb. 2022.
	
	\bibitem{Wen2022}
	H. Wen, Z. Liu, Q. Luo, C. Shi, and P. Xiao, ``Designing enhanced multi-dimensional constellations for code-domain NOMA," \emph{IEEE Wireless Communications Letters}, Jul. 2022.
	
	\bibitem{Luo2022}
	Q. Luo, H. Wen, G. Chen, Z. Liu, P. Xiao, Y. Ma, and A. Maaref, ``A novel non-coherent SCMA with massive MIMO," \emph{IEEE Wireless Communications Letters}, Aug. 2022.

	\bibitem{Lim2017}
S.-C. Lim, N. Kim, and H. Park, ``Uplink SCMA system with multiple antennas," IEEE Trans. Veh. Technol., vol. 66, no. 8, pp. 6982-6992, Aug. 2017.

	\bibitem{Md2022}
	Md. Noor-A-Rahim, Z. Liu, H. Lee, M. Omar Khyam, J. He, D. Pesch, K. Moessner, W. Saad, and H. V. Poor, ``6G for vehicle-to-everything (V2X) communications: enabling technologies, challenges, and opportunities," \emph{Proceedings of IEEE}, vol. 110, no. 6, pp. 712-734, Jun. 2022.
	
	\bibitem{He2014}
	R. He, A. F. Molisch, F. Tufvesson, Z. Zhong, B. Ai and T. Zhang, ``Vehicle-to-vehicle propagation models with large vehicle obstructions," \emph{IEEE Transactions on Intelligent Transportation Systems}, vol. 15, no. 5, pp. 2237-2248, Oct. 2014.


	\bibitem{Belmekki2020}
	B. E. Y. Belmekki, A. Hamza, and B. Escrig, ``On the outage probability of vehicular communications at intersections over Nakagami-m fading channels," \emph{2020 IEEE 91st Vehicular Technology Conference (VTC2020-Spring)}, 2020, pp. 1-5.
	
	\bibitem{Cui2004}
	T. Cui and C. Tellambura, ``An efficient generalized sphere decoder for rank-deficient MIMO systems," \emph{IEEE 60th Vehicular Technology Conference, 2004. VTC2004-Fall. 2004}, 2004, pp. 3689-3693 Vol. 5.
	
	\bibitem{Luo2021}
	Q. Luo, P. Gao, Z. Liu, L. Xia, Z. Mheich, P. Xiao, and A. Maaref, “An Error Rate Comparison of
	Power Domain Non-Orthogonal Multiple Access and Sparse Code Multiple Access," \emph{IEEE Open
	Journal of the Communications Society}, vol. 2, pp. 500-511, Mar. 2021.
	
\end{thebibliography}
\end{document}